\documentclass[10pt,twoside]{article}
\usepackage[dvips]{graphics,graphicx,color}
\usepackage{a4,indentfirst,latexsym,amsfonts,amsmath,amssymb,array,subfigure}
\usepackage{wrapfig,delarray,eucal}
\usepackage{tabularx,lscape,xr}
\usepackage[T1]{fontenc}

\title{THE TWO-BAND THEORY OF SUPERCONDUCTIVITY TURNS 50}
\author{V. B\^{a}rsan\thanks{IFIN-HH and FHH, 077125 Magurele-Bucharest, vbarsan@theory.nipne.ro;},\quad
G. Ciobanu\thanks{University of Bucharest, Dept. of Physics, PO Box MG11, 077125 Magurele - Bucharest; gheorghe.ciobanu@g.unibuc.ro.}}
\begin{document}
\maketitle

\begin{abstract}

The publication of a revised edition of a monograph devoted to the two-band theory of superconductivity provides us the opportunity of a short comment %%@
on two classical papers, published independently some 50 years ago, where the generalization of the BCS theory for metals with overlapping energy %%@
bands at the Fermi level has been proposed. The article includes Moskalenko's recollections on some aspects of scientific life in the Soviet Union, in %%@
the '50s.

\end{abstract}

\section{Introduction}

The revised edition of a monograph devoted to the two-band theory of superconductivity \cite{1}, published to honor the  $80^\text{th}$ birthday %%@
celebration of Professor Vsevolod Moskalenko (September 26, 2008), provides us the opportunity of some comments concerning two classical papers, %%@
published some 50 years ago, devoted to the generalization of BCS theory for metals with overlapping energy bands at the Fermi level. The first one %%@
was submitted to a Russian journal, \textit{Fizica Metallov i Metallovedenia}, and was published in October 15, 1959 \cite{2}; its English %%@
translation, in \textit{Physics of Metals and Metallography} \textbf{8}, 25, \cite{3}, \cite{4} was issued in June 1961. The second one, of H. Suhl, %%@
B.T. Matthias and L.R. Walker, was published in \textit{Physical Review Letters} \textbf{3}, 552, in December 15, 1959 \cite{5}. So, these papers have %%@
been written not only independently, but also in two countries separated by the Iron Curtain, in the years of the Cold War: one of them, in USSR; the %%@
other one, in USA.

The main goal of this article, which contains two parts, is to present "the untold story" of Moskalenko's paper. The first part provides a short %%@
analysis of the scientific content of the two papers. The second part is devoted to some of Moskalenko's recollections describing the scientific life %%@
in Soviet Union, in the '50s.

\section{The beginning of the two-band theory of superconductivity}

The two-band theory of superconductivity has been created, independently, in two papers, authored by Soviet and American Physicists, as already said %%@
in the Introduction. We shall give here a short description of these papers.

The starting point of Moskalenko's paper \cite{2}  is the remark that the metals which display a superconducting transition have overlapping energy %%@
bands, and this fact has not been taken into consideration neither by BCS, nor by Bogoliubov theories. So, the Hamiltonian describing the metal has a %%@
free term, with contributions from electrons belonging to two bands (the chemical potential being renormalized by interaction), and an interaction %%@
one. This last term contains a phonon-mediated intra-band, and also inter-band, electron-electron interaction. The fermionic operators entering in %%@
this Hamiltonian undergo a Bogoliubov transform, and, for the new Hamiltonian, a diagrammatic method, due to Bloch and de Dominicis, is used in order %%@
to calculate the thermodynamic potential, in the first order in the interacting potential. A compensation theorem, similar to that given by Zubarev %%@
and Tserkovnikov, is used.

The quasi-particle energy and the gaps in energy bands are obtained, and also the critical temperature . If the intra-band interaction is taken into %%@
account, both subsystems of electrons (belonging to both bands) undergo the superconducting transition at the same temperature. If the inter-band %%@
interaction is neglected, each subsystem has its own critical temperature; so, lowering the temperature, the subsystem with highest critical %%@
temperature undergoes the transition first; after further cooling, the second sub-system undergoes the transition too, at a lower critical %%@
temperature.

The jump in specific heat, the critical magnetic field and the loss in energy due to the superconducting transition are calculated.

The band with highest density of states near Fermi surface plays the dominant role in determination of the parameters of the superconducting state. %%@
The conditions to be fulfilled for occurrence of superconducting state are less rigid than in BCS case. The physics of the problem is obtained for any %%@
energy, between 0 and $T_c$.

The basic ideas of this paper have been developed by Moskalenko and his co-workers at the Institute of Applied Physics of the Academy of Science of %%@
Moldovan SSR (after 1991, the Academy of Science of the Republic of Moldova); for a review see the aforementioned monograph.

A new English translation of Moskalenko's paper is available \cite{4}. This paper is interesting both for historical reasons, and for its recent %%@
applications. It (and its developments, \cite{1}) describes the superconductivity of heavy fermion compounds CeTIn$_5$ (T=In, Rh, Co), which become %%@
superconductors at high pressure; this effect has been discovered in 2000 \cite{6}, \cite{7}. Another interesting situation is provided by a simple %%@
binary compound, MgB$_2$. The discovery of superconductivity in MgB$_2$ at $T_c = 39$ K, almost twice the critical temperature of other simple %%@
intermetallic compounds, was an "unexpected gift" for low-temperature physics. It is, clearly, a two-band superconductor, the ratio of the two gaps %%@
being 2.6 \cite{9}. Its behavior is very well described by Moskalenko's model; the compound is intensively studied, both in 3D- \cite{10}, \cite{11}, %%@
\cite{12}, \cite{13} and 2D \cite{14}, \cite{15}, \cite{16}.

In the paper of Suhl et al. \cite{5}, the electronic Hamiltonian is essentially the same; the authors make a similar Bogoliubov transform, and from %%@
the diagonalization conditions, they obtain the critical temperature and the values of energy gaps. No other physical results (jump in specific heat, %%@
critical magnetic field, calculation of thermodynamic potential) are given. However, due to the delay in the publication of Moskalenko's paper, the %%@
Western scientific community perceived the paper of Suhl et al. as the first one proposing the two-band theory of superconductivity.

\section{Recollections of a former Bogoliubov's PhD student:\\ excerpts from a discussion with Acad. Vsevolod A.
Moskalenko}

\vspace{2mm}
\textit{How did you become one of Bogoliubov's PhD students?}
\vspace{2mm}

\textbf{Acad. Vsevolod Moskalenko:} In 1950, in the \textit{Ukrainian Journal of Mathematics}, it was published the paper \textit{The adiabatic theory %%@
of interaction of an elementary particle with a quantum field}, by Bogoliubov, devoted to the polaron problem. At that time, this was a very modern %%@
problem. It was initiated by Landau and developed by Pekar. Using Landau's ideas, Pekar had shown that the elementary particle, I mean the electron, %%@
is polarizing the atoms around him and, is this way, "is digging its well", producing an attractive potential. So, the electrons become dressed %%@
particles, with a fluctuating movement through the crystal. However, the problem was put incorrectly, in the sense that the translational and %%@
fluctuational movements were not properly treated. Bogoliubov formulated a new perturbational approach. Being a man of high mathematical culture, he %%@
developed a very elegant method, which impressed me very much. After having red the article, I decided that I must absolutely find this man.

\vspace{2mm}
\textit{What were you doing at that time?}
\vspace{2mm}

\textbf{AVM:} I was a student in the last grade, the $5^\text{th}$, of the Faculty of Physics at Chi\c{s}in\u{a}u University. This was the first %%@
series of students.

\vspace{2mm}
\textit{So, you started the faculty in ....}
\vspace{2mm}

\textbf{AVM}: \ldots in 1946.

\vspace{2mm}
\emph{It was a very hard time \ldots}
\vspace{2mm}

\textbf{AVM}: It was, indeed. It was  a terrible famine, and I was officially registered as a dystrophic, with the right of receiving a bowl of cereal %%@
soup, daily. I was extremely weakened, and I was not able to walk by myself, it
was a colleague of mine who was helping me to go to the refectory. The studies were free, in general, but myself, and my brother, we had to pay the %%@
taxes. This was because we have lived "under bourgeois regime" and "under fascist occupation" (in Romania; a part of ourdays Republic of Moldova had %%@
belonged to Romania, between 1918-1940 and 1941-1944, editor's note). Our father had been deported 1940, and he died in Gulag; so, we were the sons of %%@
an "enemy of the people". This is why me and my brother were obliged to work during our studies, as laboratory assistants in the University. We could %%@
not attend the courses, we could not properly prepare the exams .... it was a very hard time, indeed.

\vspace{2mm}
\textit{So, how did you reach Bogoliubov?}
\vspace{2mm}

\textbf{AVM:} In 1951, I have decided to go to Moscow, to find him. I went to "Steklov Institute", but I could not find Bogoliubov; however I met his %%@
coworker, Tyablikov, who was a very kind and helpful person.

\vspace{2mm}
\textit{You did not write to him, to announce your visit?}
\vspace{2mm}

\textbf{AVM:} No .... I was not daring .... I went to "Steklov" every year, but only 1956 I could find Bogoliubov. It happened to be a seminar; I %%@
attended the seminar, and, after that, Bogoliubov asked me: "please, what would you like to do?" I responded that I want to study Feynman's methods; I %%@
said "I know to tackle products of two operators
(bilinear Hamiltonians, editor's note), but here (in "Steklov", editor's note), you should know how to tackle products
of four operators, too .... and I would like to learn this issue ...." In the room were also Tyablikov, Zubarev,
Tserkovnikov, Vladimirov, .... When I finished, they remained silent. "So, what shall we do with him?" - asked
Bogoliubov; they remained silent again. Then, Bogoliubov took a sheet of paper and wrote down: "Accepted for PhD,
Bogoliubov".

\vspace{2mm}
\textit{Did you remain in Moscow?}
\vspace{2mm}

AVM: No, I had to come back to Chisin\u{a}u. I tried to prepare the papers in order to get a PhD scholarship. But it
was impossible to speak to the rector; his secretary did not allow me to enter, she was treating us like zeros, like no one. However, when they found %%@
out that I was accepted by Bogoliubov, the situation changed. I was enrolled as a PhD student, but only for one year. My colleagues without "political %%@
problems" were accepted for three years.

\vspace{2mm}
\textit{When did you start the PhD preparation?}
\vspace{2mm}

\textbf{AVM:} In September 1957. I met Bogoliubov in his office at "Steklov", and I told him: "Nikolai Nikolaevich, I arrived. "And he responded, %%@
"Very well. From now onwards, you will speak to me in the language of diagrams." And I did not understand what this could mean ....

\vspace{2mm}
\textit{How did you start?}
\vspace{2mm}

\textbf{AVM:} At the beginning, I was finishing my papers, which I had started at Chi\c{s}in\u{a}u. I was in a tight competition with Hacken, who was %%@
studying the same problem - the exciton. My first paper in JETF was devoted to the theory of excitons (JETF, \textbf{30}, 959 (1956), editor's note); %%@
this paper was included by Hacken in his review on excitons. I was afraid that Hacken will finish the work before me, and every morning, when I was %%@
arriving in the
library, I was searching if Hacken's paper appeared or not ... Finally, I finished the article and I wanted to show it to Bogoliubov. But he was %%@
surrounded by a lot of people like a polaron, and I did not dare to disturb him \ldots

\vspace{2mm}
\textit{So, you did not show him your article?}
\vspace{2mm}

\textbf{AVM:} I did, but a little bit later \ldots the institute has a large stair, with windows, and I could see him when he was descending, to go %%@
home. So, I waited for him on stairs and I have shown him the paper. He took the paper and red it quickly and told me, showing an equation: "Look, %%@
here the dead dog is buried." And nothing more. And I was thinking and working again, and after three months I could find the error \ldots

\vspace{2mm}
\textit{So, he did not use to explain much to you \ldots}
\vspace{2mm}

\textbf{AVM:} Not much ... when some colleagues tried to ask him how to proceed, he responded: "Which would be your benefit if I shall tell you what %%@
to do?" However, for me, only the fact that I could stay around such a man was a great happiness. I was avoiding, as much as possible, to consume his %%@
time. This is way I was using to wait for him on stairs and to have there short discussions. But he was a democrat. I shall give you an example. %%@
During my PhD, we, the
youngest researchers, were working it the same room. And Bogoliubov, when he was leaving the Institute to go home, used to come to each of us, to %%@
shake the hands and to say: "I am greeting you, I am greeting you ..." Of course, I was in the most distant place ... and once it happened  that he %%@
passed near all my colleagues, and was about to get out, but he noticed that he did not shake the hands with me; then, he came back, went to my desk %%@
and said to me "I am greeting you". I shall keep in my heart forever this "I am greeting you". Especially when you compare with the atmosphere in %%@
Chi\c{s}in\u{a}u, where if you go to rector's office, nobody pays any attention to you ... Bogoliubov was a very direct man, sometimes he was kidding, %%@
but very seldom. He was very cultured, knowing very well Russian literature; he used to say, by heart, long quotations.

He was speaking very wittily, and often I had to ask somebody to explain to me what he really meant.

So, in my life, I had the chance of knowing a giant, Bogoliubov, and also a man of highest moral attitude - Tyablikov.
Tyablikov used to say the truth right in face; this was very rare. He was elegant, handsome, slim. He was very kind, very helpful, but he was helping %%@
people discretely, without speaking about this; he was trying to help you, and not to
let you know about this. I learned a lot of things from him.

\vspace{2mm}
\textit{But coming back to the previous question, how did you start to work in superconductivity?}
\vspace{2mm}

\textbf{AVM:} I shall tell you. I the autumn of '57, everybody was discussing about superconductivity. Previously, in every year, in '55, in '56, a %%@
theory of superconductivity was published, and all of them were wrong. The theories of Fr\"{o}hlich, of Schaffroth, of Blatt.... when the preprint of %%@
BCS article arrived, the people said "look, a new wrong
theory", and they throw it away .... Of course, Cooper's paper was well known, and I personally heard Landau saying that something more %%@
incomprehensible than this paper, he never had seen. It was a great excitement, with a lot of discussions and seminars: at Lomonosov University; at %%@
Steklov, Bogoliubov's seminar; at the Institute of Physics of the Academy of Sciences, Ginzburg's seminar; of course Landau's seminar and the best %%@
rated officially, Kapitza's seminar. People like Landau, Fock, Pomeranchuk, Gorkov, Abrikosov were attending such seminars. Bogoliubov presented his %%@
theory, including the Fr\"{o}hlich interaction. BCS theory took into consideration only the electron gas, the interaction with
phonons being replaced with an effective interaction between electrons; but Bogoliubov considered the electron gas, the phonon gas, and the %%@
Fr\"{o}hlich interaction. In this atmosphere, I decided to change the domain of  my
researches and to study the superconductivity. My colleagues told me, "what are you doing, you are here for only one year, and you drop your familiar %%@
themes and start something absolutely new?" It was not a wise decision, but it was impossible to me to act otherwise.

\vspace{2mm}
\textit{So, how did you really start?}
\vspace{2mm}

\textbf{AVM:} It was in May '58. Bogoliubov and Tyablikov were discussing near the blackboard, and Bogoliubov was saying: "You see, these theories %%@
(BSC and Bogoliubov's, editor's note) are made for some ideal gases; you cannot see the dependence on real metals, on their parameters; we should do a %%@
theory for real metals." I was sitting somewhere in
the room, listening. It was the time when Tyablikov was preparing himself for holidays. He used to go in June, alone,
in Siberia, just with a gun, in those frightening forests ...

\vspace{2mm}
\textit{In the taiga?}
\vspace{2mm}

\textbf{AVM:} In the taiga, where the villages are 100 km away each other, and the bears smell your traces and, if you
see a man, you must ask yourself if he will shut you, or you should shut him ... And Tyablikov tells me, "Vsevolod Anatolievich, come with me in %%@
taiga, I shall pay the ticket for you, we shall go together in Kamenaya - Tonguska", where the meteorite had fallen ... "Thank you so much Sergei %%@
Vladimirovich, but I am here for only one year, so how could I go for two months in Siberian forests? So much the more that my wife with my two girls %%@
will arrive, and I must be here, to welcome them. "Very well, said Tyablikov, you will take up this problem" (the superconductivity of real metals, %%@
editor's note).

\vspace{2mm}
\textit{You did not regret not to join Tyablikov?}
\vspace{2mm}

\textbf{AVM:} You know, the science was the priority. So, I remained in Moscow; it was very hot, and I did not know German, and I had to read the %%@
theory of metals, of Sommerfeld, and "Zeitschrift", and "Annalen der Physik"...

\vspace{2mm}
\textit{They were not translated into Russian?}
\vspace{2mm}

\textbf{AVM:} No ... and no textbook, which are now so abundant. I was working hardly, it was very hot, and I was not understanding ... I was trying %%@
to use the Umklapp processes, but I could not achieve anything.... and September was coming, on September 15 Tyablikov will be back, and how can I %%@
meet him with empty hands, when he told me: "take up the problem"? You know, such an imperative is terrible. I was in a very tense state.... and I put %%@
together all may forces, and something more, and I said to myself: I must do something..... And in this way, the two-band theory appeared. I got the %%@
idea, and in two weeks I wrote the paper. When Tyablikov came back, the paper was ready.

\vspace{2mm}
\textit{Where did you submit the paper?}
\vspace{2mm}

\textbf{AVM:} I submitted in to JETF (Journal of Experimental and Theoretical Phy\-sics, the most prestigious Soviet journal of physics, editor's %%@
note); from September till October, it was rejected.

\vspace{2mm}
\textit{Why?}
\vspace{2mm}

\textbf{AVM:} "It does not present interest".

\vspace{2mm}
\textit{Who said this?}
\vspace{2mm}

\textbf{AVM:} The referee was secret, of course.

\vspace{2mm}
\textit{It was not possible to ask for another referee? Even if the article was considered to be very good, by people like Tyablikov and Bogoliubov?}
\vspace{2mm}

\textbf{AVM:} No way. So we decided to send the paper to Sverdlovsk, at "Physics of metals and metallography", whose chief editor was Vonsovskii (at %%@
JETF, the chief editor was Kapitza, and the deputy was Evgueni Lifshitz, editor's note). The paper was submitted in October 1958, and published in %%@
October 1959. But Suhl, Mathias and Walker submitted their paper to Physical Review Letters on November 15, 1959, and it was published on December 15, %%@
1959.

\vspace{5mm}
\textit{As already mentioned, the English translation of that volume of the "Physics of metals and metallography" appeared in June 1961, having a %%@
quite limited circulation. This is why, for Western scholars, the paper proposing the two-band theory was the paper of Suhl et al., and Moskalenko's %%@
paper passed largely unnoticed. (editor's note).}


\begin{thebibliography}{99}

\bibitem{1} V. A. Moskalenko, L. Z. Kon, M. E. Palistrant: \textit{Theory of multi-band superconductors} (in Romanian), Editura Tehnica, Bucure\c{s}ti %%@
(2008); English translation (see \cite{4}): http://www.theory.nipne.ro/vbarsan/ebooks/Mosk2008
\bibitem{2} V. A. Moskalenko, Fiz.Met.Metallov. \textbf{8}, 503 (1959)
\bibitem{3} V. A. Moskalenko, Phys.Met.Metallogr. \textbf{8}, 25 (issued in 1991)
\bibitem{4} http://www.theory.nipne.ro/~vbarsan/docs/publications.pdf
\bibitem{5} H. Suhl, B. T. Matthias, L. R. Walker, Phys.Rev.Lett. \textbf{3}, 552 (1959)
\bibitem{6} H. Heeger et al., Phys.Rev.Lett. \textbf{84}, 4986 (2000)
\bibitem{7} A. Bussman-Holder et al., Europhys.Lett. Phys.Rev.Lett. \textbf{75}, 308 (2006)
\bibitem{8} J. Nagamatsu et al., Nature (London) \textbf{63}, 401 (2001)
\bibitem{9} R. S. Gonelli et al., Supercond.Sci.Tech. \textbf{16}, 171 (2003)
\bibitem{10} P. Udomsamuthirun et al., physica status solidi (b) \textbf{230}, 591 (2003)
\bibitem{11} J. N. Askerade et al., J.Phys.Soc.Jap. \textbf{71}, 1637 (2002), Physica Scripta \textbf{69}, 234 (2004)
\bibitem{12} J. Cortus et al., Phys.Rev.Lett. \textbf{94}, 027002 (2005); Phys.Rev. \textbf{B72}, 024504 (2005); Physica \textbf{C456}, 54 (2007)
\bibitem{13} Gh. Aldica et al., J.Optoel.Adv.Mat. \textbf{9}, 1742 (2007); \textbf{10}, 929 (2008)
\bibitem{14} W. N. Kang et al., Science \textbf{292}, 1521 (2001)
\bibitem{15} X. H. Zeng et al., Nature Mater. \textbf{1}, 35 (2002)
\bibitem{16} L. Petaccia et al., New J.Phys. \textbf{8}, 12 (2006)

\end{thebibliography}
\end{document}